# High temperature superconducting FeSe films on $SrTiO_3$ substrates

Yi Sun[1†], Wenhao Zhang[2,3†], Ying Xing[1], Fangsen Li[2,3], Yanfei Zhao[1], Zhengcai Xia[5], Lili Wang[2,3,4*], Xucun Ma[2,3,4], Qi-Kun Xue[2,4], Jian Wang[1,4*]

[1] *International Center for Quantum Materials, School of Physics, Peking University, Beijing 100871, China*

[2] *State Key Laboratory of Low-Dimensional Quantum Physics, Department of Physics, Tsinghua University, Beijing 100084, China*

[3] *Institute of Physics, Chinese Academy of Sciences, Beijing 100190, China*

[4] *Collaborative Innovation Center of Quantum Matter, Beijing 100871, China*

[5] *Wuhan National High Magnetic Field Center, Huazhong University of Science and Technology, 1037 Luoyu road, Wuhan 430071, China*

Interface enhanced superconductivity at two dimensional limit has become one of most intriguing research directions in condensed matter physics. Here, we report the superconducting properties of ultra-thin FeSe films with the thickness of one unit cell (1-UC) grown on conductive and insulating $SrTiO_3$ (STO) substrates. For the 1-UC FeSe on conductive STO substrate (Nb-STO), the magnetization *versus* temperature (*M-T*) measurement shows a diamagnetic signal at 85 K, suggesting the possibility of superconductivity appears at this high temperature. For the FeSe films on insulating STO substrate, systematic transport measurements were carried out and the sheet resistance of FeSe films exhibits Arrhenius TAFF behavior with a crossover from a single-vortex pinning region to a collective creep region. More intriguing, sign reversal of Hall resistance with temperature is observed, demonstrating a crossover from hole conduction to electron conduction above $T_C$ in 1-UC FeSe films.

[†]These authors contributed equally to this work.

[*]Correspondence and requests for materials should be addressed to J.W.(jianwangphysics@pku.edu.cn) or L.W.(liliwang@mail.tsinghua.edu.cn)

## INTRODUCTION

Iron-based superconductors has triggered great interest [1] because of the high transition temperature [2,3], ultra-high critical magnetic field [4-6] and potential applications as a group of high $T_C$ superconductors[7,8]. FeSe, with the simplest structure and less toxicity in iron-based superconductors (among all five families), has become one of attractive materials but the $T_C$ is relatively low in bulk state (～9 K). In the meanwhile, heterostructure based interface engineering has been proved an effective method for raising $T_C$ due to the enhancement of electron-phonon coupling [9] or epitaxial strain [10]. In previous work, *in situ* scanning tunneling microscopy/spectroscopy (STM/STS) [11] and angle resolved photoemission spectroscopy (ARPES) [12-14] detections on 1-UC thick FeSe films on Nb-doped STO (conductive STO) substrates revealed a superconducting energy gap as large as 20 meV and above 15 meV (closing at a temperature of 65±5 K) respectively, indicating a possible $T_C$ higher than 60 K. Following, the $T_C$ above 40 K in 1-UC FeSe films on insulating STO has been demonstrated by direct transport measurements and Meissner effect [7]. However, direct evidence of $T_C$ above 60 K for FeSe films on conductive substrates and systematic transport studies of FeSe films on insulating substrates are still absent.

## EXPERIMENTAL

Since previous STM study indicates that the second UC of FeSe films grown on STO substrates shows semiconducting behavior and only the first UC FeSe is superconducting, carefully comparison of the transport properties of high quality 1-UC and 2-UC FeSe films on insulating STO substrates becomes necessary. In this paper, we mainly report the electronic transport and diamagnetic results from five typical FeSe films grown by molecular beam epitaxy (MBE) system. Sample 1 is 1-UC FeSe on conductive STO substrate, samples No.2 and No.3 are 1-UC FeSe films on insulating STO substrates, while samples No.4 and No.5 are 2-UC FeSe films on insulating STO substrates. The 10-UC FeTe protection layers are deposited before the *ex situ* measurements of the films [11]. All FeSe films are 1.5 mm wide and 8 mm long. The distance between the voltage electrodes for the measurement is 1.8 mm for sample No.2, 2.25 mm for sample No.3, 2.0 mm for sample No.4 and 2.75 mm for sample No.5. As for sample No.1, the diamagnetic measurement exhibits that the onset of magnetization drop starts around 85 K, indicating a possible superconductivity up to this high temperature. For samples No. 2-5, thermally activated flux flow (TAFF) behavior and Hall effect are carefully studied. It is found that single-vortex pinning dominates vortex dynamics in low magnetic field region, whereas collective creep becomes important at higher magnetic fields. More interestingly, sign reversal behavior of Hall coefficient ($R_H$) with temperature is observed.

## RESULTS AND DISCUSSIONS

It is worthy to mention that the observed $T_C$ of 1-UC FeSe on insulating substrate by *ex situ* transport measurements is obvious lower than estimated $T_C$ of 1-UC FeSe on conductive substrate by STM and ARPES studies [11-14]. One of major difference of the two kinds of 1-UC samples is the conductance of the substrates. In order to see if higher $T_C$ of 1-UC FeSe can be revealed by direct measurements other than energy gap detection, we did magnetization experiments for 1-UC FeSe on conductive STO substrate (sample No.1) in a magnetic property measurement system (MPMS-SQUID-VSM). DC magnetization as a function of temperature (*M-T*) during both zero field cooling (ZFC) and field cooling (FC) at 1000 Oe of sample No.1 is shown in Fig.1(a). The *M-T* curves exhibit a diamagnetic drop around ~ 85 K, which offers a hint that the superconductivity in 1-UC FeSe may exist up to 85 K. Please notice that there is no such a drop for the non-superconducting 10 UC FeTe layer on conductive STO substrate as shown in the inset of Fig. 1(a) for comparison. Following, a much sharper drop indicating characteristic of Meissner effect appears at around 21 K, which is close to the $T_c^{zero}$ obtained by transport measurement for 1-UC FeSe films on insulating STO. After subtracting the influence of STO substrate and FeTe protection layer, the *M-T* curves of 1-UC FeSe film exhibit more apparent diamagnetic drop at 85 K with decreasing temperature (shown in Fig. 1(b)).

The electronic transport measurements were carried out in a physical property measurement system with the magnetic field up to 16 T (PPMS-16T). The obtained superconducting parameters of the four measured samples are summarized in Table 1. Figure 2 shows the transport results of sample No.2, one typical 1-UC FeSe film grown on insulating STO substrate covered by non-superconducting FeTe protection layers with an excitation current of 500 nA. Fig. 2(a) shows the sheet resistance of the sample as a function of temperature $R_{sq}(T)$ at zero magnetic field ($\mu_0 H$). The resistance begins to drop at about 54.5 K. By extrapolating both the normal resistance and the superconducting transition curves, we obtain the onset $T_c^{onset}$= 40.2 K and the resistance drops completely to zero at 23.5 K ($T_c^{zero}$). The zero resistance is defined when the measured voltage is within the instrumental resolution ±20 nV. Figure 2(b) shows $R_{sq}(T)$curves at different perpendicular magnetic fields ($\mu_0 H$) up to 16 T from 2 K to 60 K. The resistive transition becomes broader and shifts to lower temperatures with increasing magnetic field, characteristic of superconducting transition in thin films.

We know in the thermally activated flux flow (TAFF) of vortex region, the ln$\rho$− $1/T$ can be described by Arrhenius relation [15,16]

$$\rho(T,H) =\rho_0(H) \exp[- U_0(H)/T] \qquad (1)$$

where $\rho_0$ is a temperature independent constant and $U_0(H)$ is the activation energy of the flux flow. Thus, ln$\rho(T,H)$ versus $1/T$ should be linear in the TAFF region. As shown in Fig. 2(c), the experimental data in Fig. 2(b) can be well fitted by the Arrhenius relation (solid lines). The fitting lines obtained from ln$\rho(T,H)$ versus $1/T$ at different magnetic fields can be well extrapolated to the same temperature, $T_m$= 38 K, which is close to the value of $T_C^{onset}$. In addition, the activation energy $U_0$ for different magnetic field can be obtained by the slope of the solid fitting lines. As shown in Fig. 2(d), the $U_0(H)$ shows a magnetic field dependent power law relation,

$$U_0(H) \propto H^{-\alpha} \qquad (2)$$

with $\alpha$=0.14 ($\mu_0 H$ < 3.4 T) and 0.60 ($\mu_0 H$ > 3.4 T), respectively.

For the 2-UC FeSe film grown on insulating STO substrate covered by non-superconducting FeTe protection layer (sample No.4), Fig. 3(a) shows the sheet resistance as a function of temperature $R_{sq}(T)$ at zero magnetic field ($\mu_0 H$). By extrapolating both the normal resistance and the superconducting transition curves, we obtain the onset $T_c^{onset}$= 43 K and the resistance drops completely to zero at 22.5 K ($T_c^{zero}$), which are almost same with those of 1-UC FeSe (sample No.2). Figure 3(b) shows the $R_{sq}$ of sample No.4 as a function of temperature, $R_{sq}(T)$, at different perpendicular magnetic fields ($\mu_0 H$). The $R_{sq}(T)$ curves at different magnetic fields can also be fitted by the Arrhenius relation (solid lines) well as shown in Fig. 3(c). The fitting lines obtained from ln$\rho(T,H)$ versus $1/T$ at different magnetic fields cross to one point at about $T_m$= 36 K, which is consistent with the $T_C$ of the sample. The activation energy $U_0$ varies with different magnetic field can be achieved by the slope of the solid fitting lines as shown in Fig. 3(d), $\alpha$=0.125 ($\mu_0 H <$ 3.7 T) and 0.79 ($\mu_0 H >$ 3.7 T) respectively by utilizing Eq. (2). These values obtained from the two samples are close. At the temperatures not far from $T_C$, the weak power law decreases of $U_0(H)$ in low magnetic fields for both samples implies that single-vortex pinning dominates in this region, followed by a quicker decrease of $U_0(H)$ at high field where a crossover to collective flux creep regime occurs. The similar behaviors are observed in iron-based superconductors, such as FeSe [17,18], FeTeSe [19] and Nd(O,F)FeAs [20] crystals. The exponent $\alpha$ = 0.5 and 1 corresponds to a planar-defect pinning and a point-defect pinning in high $T_C$ superconductors respectively [21]. For our FeSe ultrathin films, the fitted values obtained at high magnetic field vary between 0.5 and 1, suggesting that the pinning centers may be mixed with point and planar defects. The cross-over magnetic field is about 3 T for bulk FeSe, 2 T for FeTeSe, and 3 T for Nd(O,F)FeAs, which are almost in the same level with our observation in ultrathin FeSe films. For cuprates, the range of the cross-over magnetic field is from 0.8 T - 5.5 T, where the exact value depends on the sample situations, such as defects and boundaries.

The Hall resistance ($R_{xy}$) of FeSe films is measured by sweeping the magnetic field at a fixed temperature. The temperature stabilization is better than 0.1%. The distance between the Hall voltage electrodes is about 1.5 mm for all measured samples. Figures 4(a)-4(b) show the data of $R_{xy}$ vs magnetic field ($R_{xy}(H)$) at different temperatures from 40 K to 150 K of samples No.2 and No.4, which exhibit good linear relation. In order to subtract the influence of FeTe protection layer, transport properties of reference sample (10 UC FeTe grown on insulating STO substrate) were studied. The sheet resistance of the reference sample varies with temperature ($R_{sq}(T)$) under different magnetic fields (0 T, 9 T and -9 T) are exhibited in Fig. 4(c). Figure 4(d) shows the $R_{xy}(H)$ curves of the reference sample at different temperatures. The Hall conductance of pure FeSe films (removed the influence of FeTe layers) can be calculated by $\sigma_{xy(FeSe)} = \sigma_{xy(sample)} - \sigma_{xy(FeTe)}$ ($\sigma_{xy} = -\frac{R_{sq(xy)}}{R^2_{sq(xx)}+R^2_{sq(xy)}}$). After removing the influence of FeTe protection layer, the Hall resistance ($R_{sq(xy)-FeSe}$) of the pure FeSe film sample is re-obtained by conversing from Hall conductance of FeSe films, $R_{sq(xy)-FeSe} = -\frac{\sigma_{xy(FeSe)}}{\sigma^2_{xx(FeSe)}+\sigma^2_{xy(FeSe)}}$ (where $\sigma_{xx} = \frac{R_{sq(xx)}}{R^2_{sq(xx)}+R^2_{sq(xy)}}$), which are plotted in Figs. 5(a) and 5(c) for sample No.2 and No.4, respectively.

In Fig. 5(b), the temperature dependence of the Hall sensitivity $\frac{R_H}{d} = \frac{V_H}{I(\mu_0 H)} = \frac{R_{xy}}{\mu_0 H}$ ($V_H$ is the Hall voltage, $d$ is the thickness of the sample) and two-dimensional carrier density ($n_S = \frac{1}{(R_H/d)\cdot q} = \frac{1}{(R_{xy}/\mu_0 B)\cdot q}$, $q$ is the charge of electron) are plotted. For a normal metal with Fermi liquid feature, the Hall coefficient is constant at different temperatures. However, it varies with temperature for multiband materials, such as $MgB_2$ [22], iron-based superconductors [23], or a sample with non-Fermi liquid behavior such as cuprate superconductors [24]. Here, an interesting phenomenon—sign reversal behavior of $\frac{R_H}{d}$ is observed in the 1-UC FeSe film (sample No.2) with temperature increasing (shown in Fig. 5(a) and 5(b)). In Fig. 5(a), the slope of the $R_{xy}(H)$ curves changes from negative to positive as the temperature is higher than 100 K. Correspondingly, the signs of $\frac{R_H}{d}$ and $n_s$ reversed at the same time as shown in Fig. 5(b). That is to say, as the temperature is lower than 100 K, the 1-UC FeSe film is electron-doped, but it is hole-doped while the temperature higher than 100 K. As for 2-UC FeSe film, Figure 5(c) shows the $R_{xy}(H)$ curves of sample No.4 at different temperatures from 40 K to 150 K. The Hall sensitivity $\frac{R_H}{d}$ for sample No.4 decays continuously with increasing temperature as shown in Fig. 5(d). The value of $\frac{R_H}{d}$ at 150 K is almost 10 times smaller than that at 40 K, exhibiting strong temperature dependence of $\frac{R_H}{d}$. This behavior is possibly induced by the multiband effect. For sample No.4, the Hall data manifest electron-doped property as increasing temperature till 150 K. Please notice that the carrier density ($n_s$) in FeSe films is pretty huge (~$10^{15}$ cm$^{-2}$).

In order to confirm that the sign reversal of $\frac{R_H}{d}$ is a universal behavior in ultrathin FeSe films, Hall resistance from another two samples (sample No.3 is 1-UC FeSe film and sample No.5 is 2-UC FeSe film) at various temperatures up to 300 K was measured. Figure 6(a) and 6(b) are the raw $R_{xy}(H)$ data from sample No.3 and No.5. The signs of $\frac{R_H}{d}$ are obviously reversed in these two samples with increasing temperature. After subtracting the influence of FeTe protection layer utilizing the same method mentioned above, the $R_{xy}$ curves of pure FeSe films at different temperatures are shown in Fig. 6(c)-6(f) with similar sign reversal behavior of $\frac{R_H}{d}$. The parameters of Hall effect for all four samples can be found in Table 2 & 3. Thus, the crossover of conduction carrier type with temperature is demonstrated to be a universal phenomenon for ultrathin FeSe films. The reversal temperature is 80 K or 150 K for 1-UC FeSe films (sample No.3 or sample No.2) but 150 K or larger for 2-UC FeSe (sample No.5 or sample No.4).

In summary, superconducting properties of ultra-thin (1-UC or 2-UC) FeSe films grown on insulating STO and conductive STO substrates were studied by transport and magnetic measurements. *M-T* observation of FeSe films on conductive STO shows a diamagnetic drop ~ 85 K, probably suggesting a high $T_C$ up to 85 K. The results from films on insulating STO reveal Arrhenius TAFF behavior with a transition from single-vortex pinning region to collective creep region. More intriguingly, the observed sign reversal of Hall coefficient above $T_C$ demonstrates a crossover from hole transport to electron conduction in ultrathin FeSe films with decreasing temperature.

**Acknowledgments**

The authors would like to thank Fa Wang, Zhili Xiao, Jiyong Yang and Mingliang Tian for fruitful discussions. This work was financially supported by National Basic Research Program of China (Grant Nos. 2013CB934600 & 2012CB921300), the National Natural Science Foundation of China (Nos. 11222434, 11174007, 91121004, 11321091 and 11374336), the Research Fund for the Doctoral Program of Higher Education (RFDP) of China and China Postdoctoral Science Foundation (No. 2011M500180 & No. 2012T50012).

**Author contributions**

J.W., L.W, X.M. and Q.-K.X. designed and coordinated the experiments; Y.S., W.Z., Y.X., F.L., Y.Z., Z.X. and J.W. carried out the experiments; Y.S. and J.W. wrote the paper.

**Additional information**

**Competing financial interests:** The authors declare no competing financial interests.

**Figure captions**

**Fig.1 Diamagnetic measurements of 1-UC FeSe films grown on conductive STO (LR STO) substrate (sample No.1). (a)** *M-T* curves of sample No.1 (LR STO/1-UC FeSe/10-FeTe) measured under a 1000 Oe magnetic field. Inset: *M-T* curves of LR STO/10-UC FeTe. **(b)** *M-T* curves of 1-UC FeSe (removing the influence of STO substrate and FeTe protection layers). The *M-T* curves subtracted the influence of substrate and protection layer show a possible superconducting signal at about 85 K, and following a sharp drop around 21 K.

**Fig.2 Transport measurement of the 1-UC FeSe film grown on insulating STO (HR STO) substrate (sample No.2). (a)** The temperature dependence of sheet resistance under zero field, showing $T_C^{onset}$= 40.2 K and $T_C^{zero}$= 23.5 K. **(b)** The temperature dependence of sheet resistance under various perpendicular magnetic fields up to 16 T, showing a typical broadened superconducting transition. **(c)** $lg\rho(T,H)$ *vs.* $1/T$ in various perpendicular magnetic fields. The corresponding solid lines are fitting results from the Arrhenius relation. **(d)** Field dependence of $U_0(H)$. The solid lines are power-law fits using $U_0(H) \sim H^{-\alpha}$. For this 1-UC FeSe film, $\alpha = 0.14$ for $\mu_0 H < 3.4$ T, and $\alpha = 0.60$ for $\mu_0 H > 3.4$ T.

**Figure 3. Transport measurement of the 2-UC FeSe film grown on HR STO substrate (sample No.4). (a)** The temperature dependence of sheet resistance under zero field, showing $T_C^{onset} = 43K$ and $T_C^{zero} = 22.5K$. **(b)** The temperature dependence of sheet resistance under various perpendicular magnetic fields. **(c)** $lg\rho(T,H)$ *vs.* $1/T$ in various perpendicular magnetic fields. The corresponding solid lines are fitting results from the Arrhenius relation. **(d)** Field dependence of $U_0(H)$. The solid lines are power-law fits using $U_0(H) \sim H^{-\alpha}$. For this 2-UC FeSe film, $\alpha = 0.125$ for $\mu_0 H < 3.7$ T, and $\alpha = 0.79$ for $\mu_0 H > 3.7$ T.

**Fig.4 Hall results of FeSe films and FeTe protection layers. (a)** Hall resistance ($R_{xy}$) varies with magnetic field — $R_{xy}(H)$ at different temperatures of sample No.2. **(b)** $R_{xy}(H)$ curves at different temperatures of sample No.4. **(c)** $R_{sq}(T)$ curves of the 10-UC FeTe protection layer under different magnetic fields (0 T, 9 T and -9 T). **(d)** $R_{xy}(H)$ curves at different temperatures of the 10-UC FeTe protection layer.

**Fig.5 Hall results of sample No.2 (1-UC FeSe films grown on HR STO) and sample No.4 (2-UC FeSe films grown on HR STO) after subtracting the influence of the protection layer. (a) & (c),** $R_{xy}$ vs magnetic field curves at different temperatures of sample No.2 and No.4 subtracting the background of protection layers, respectively. **(b) & (d),** Hall coefficient and the carrier density of sample No.2 and No.4 at different temperatures obtained from the data in (a) and (c).

**Fig.6 Hall results of sample No.3 (1-UC FeSe films grown on HR STO) and sample No.5 (2-UC FeSe films grown on HR STO). (a)** $R_{xy}(H)$ curves at different temperatures of sample No.3. **(b)** $R_{xy}(H)$ curves at different temperatures of sample No.5. **(c) & (d)** $R_{xy}(H)$ curves at different temperatures of sample No.3 and No.5 subtracting the background, respectively. **(e) & (f)** Hall coefficient and the carrier density of sample No.3 and No.5 at different temperatures after subtracting the background.

**Table 1. Summary of the parameters of four FeSe samples on insulating STO.** $T_C^{onset}$ was obtained by extrapolating both the normal resistance and the superconducting transition curves. $T_C^{onset}$ with star is the temperature at which the resistance starts to decrease. Since previous STM study[11] indicates that the second UC of FeSe films grown on STO substrates shows semiconducting behavior and only the first UC FeSe is superconducting, $J_C$ of the four samples at 2 K is calculated from $I_C$ by using the thickness of 0.55 nm (1-UC FeSe).

**Table 2. Hall coefficient, carrier density and mobility of four FeSe samples on HR STO at different temperatures.**

**Table 3. Hall coefficient, carrier density and mobility of four samples on HR STO subtracting the influence of FeTe protection layer at different temperatures.**

| $T_C^{onset}$ (K) | | | | $T_C^{zero}$ (K) | | | | $J_C$ (A/cm$^2$) | | | |
|---|---|---|---|---|---|---|---|---|---|---|---|
| 1 UC (No.2) | 1 UC (No.3) | 2 UC (No.4) | 2 UC (No.5) | 1 UC (No.2) | 1 UC (No.3) | 2 UC (No.4) | 2 UC (No.5) | 1 UC (No.2) | 1 UC (No.3) | 2 UC (No.4) | 2 UC (No.5) |
| 40.2 K<br>*54.5K | 36.0 K<br>*45.8K | 43.0 K<br>*50.0K | 40.0 K<br>*55.0K | 23.5 K | 2.3 K | 22.5 K | 16.5 K | $1.67\times10^6$ | $4.85\times10^4$ | $2.72\times10^5$ | $9.7\times10^5$ |

**Table 1**

| *T* (K) | $R_{xy}/\mu_0 B$ (Ω/T) | | | | $n_S$ ($10^{15}$ cm$^{-2}$) | | | | Mobility (cm$^2$/Vs) | | | |
|---|---|---|---|---|---|---|---|---|---|---|---|---|
| | 1 UC (No.2) | 1 UC (No.3) | 2 UC (No.4) | 2 UC (No.5) | 1 UC (No.2) | 1 UC (No.3) | 2 UC (No.4) | 2 UC (No.5) | 1 UC (No.2) | 1 UC (No.3) | 2 UC (No.4) | 2 UC (No.5) |
| 30 K | | -0.221 | | | | -2.828 | | | | 2.290 | | |
| 40 K | -1.545 | -0.220 | -1.073 | -0.845 | -0.405 | -2.836 | -0.583 | -0.740 | 52.150 | 1.971 | 30.970 | 12.382 |
| 50 K | -1.069 | -0.164 | -0.990 | -0.322 | -0.585 | -3.811 | -0.632 | -0.937 | 32.354 | 1.435 | 24.110 | 8.935 |
| 60 K | -0.732 | -0.104 | -0.917 | -0.589 | -0.854 | -6.013 | -0.681 | -1.060 | 23.446 | 0.903 | 20.752 | 7.623 |
| 80 K | -0.390 | -0.017 | -0.790 | -0.440 | -1.6 | -35.837 | -0.791 | -1.422 | 11.628 | 0.150 | 15.931 | 5.40 |
| 100 K | -0.212 | 0.062 | -0.643 | -0.335 | -2.96 | 10.158 | -0.973 | -1.865 | 5.859 | 0.523 | 11.437 | 3.900 |
| 150 K | 0.086 | 0.174 | -0.220 | -0.017 | 7.29 | 3.601 | -2.84 | -35.755 | 2.022 | 1.423 | 2.746 | 0.173 |
| 200 K | | 0.214 | | 0.137 | | 2.921 | | 4.552 | | 1.737 | | 1.223 |
| 300 K | | 0.188 | | 0.197 | | 3.320 | | 3.172 | | 1.583 | | 1.720 |

**Table 2 (with FeTe protection layer)**

| *T* (K) | $R_{xy}/\mu_0 B$ (Ω/T) | | | | $n_S$ ($10^{15}$ cm$^{-2}$) | | | | Mobility (cm$^2$/Vs) | | | |
|---|---|---|---|---|---|---|---|---|---|---|---|---|
| | 1 UC (No.2) | 1 UC (No.3) | 2 UC (No.4) | 2 UC (No.5) | 1 UC (No.2) | 1 UC (No.3) | 2 UC (No.4) | 2 UC (No.5) | 1 UC (No.2) | 1 UC (No.3) | 2 UC (No.4) | 2 UC (No.5) |
| 50 K | -1.305 | -0.315 | -1.287 | -0.897 | -0.179 | -1.985 | -0.486 | -0.697 | 39.516 | 3.719 | 31.343 | 19.827 |
| 100 K | -0.286 | 0.072 | -0.984 | -0.473 | -2.183 | 8.648 | -0.635 | -1.322 | 7.138 | 0.830 | 17.516 | 9.072 |
| 150 K | 0.081 | 0.257 | -0.463 | -0.073 | 7.719 | 2.432 | -1.350 | -8.522 | 1.719 | 2.432 | 5.776 | 1.197 |
| 200 K | | | | 0.214 | | | | 2.917 | | | | 3.149 |
| 300 K | | | | 0.329 | | | | 1.897 | | | | 4.744 |

**Table 3 (subtracting FeTe affection)**

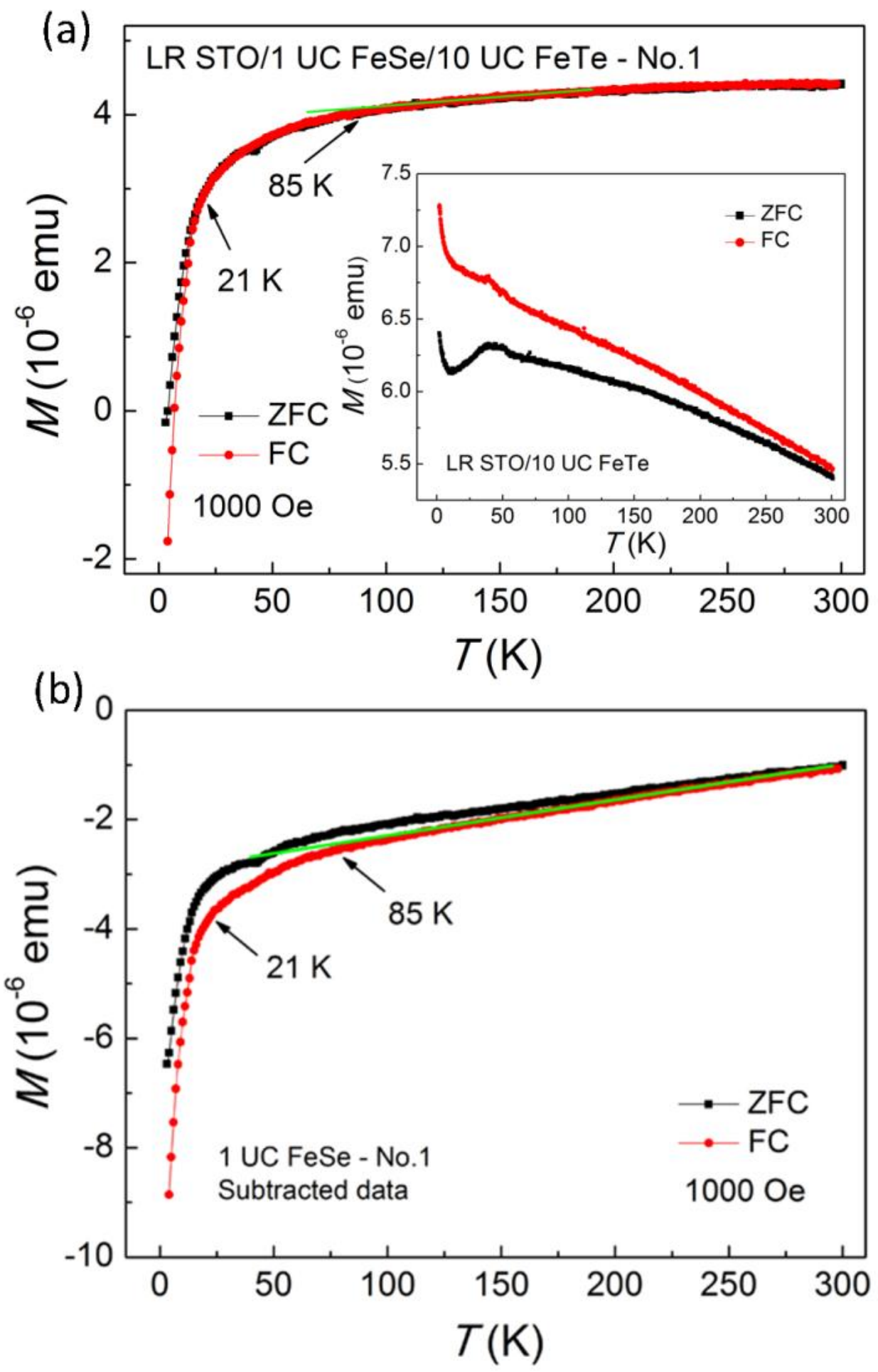
(a)
LR STO/1 UC FeSe/10 UC FeTe - No.1
85 K
21 K
ZFC
FC
1000 Oe
M (10^-6 emu)
T (K)
LR STO/10 UC FeTe
(b)
85 K
21 K
ZFC
FC
1 UC FeSe - No.1
Subtracted data
1000 Oe

**Figure 1**

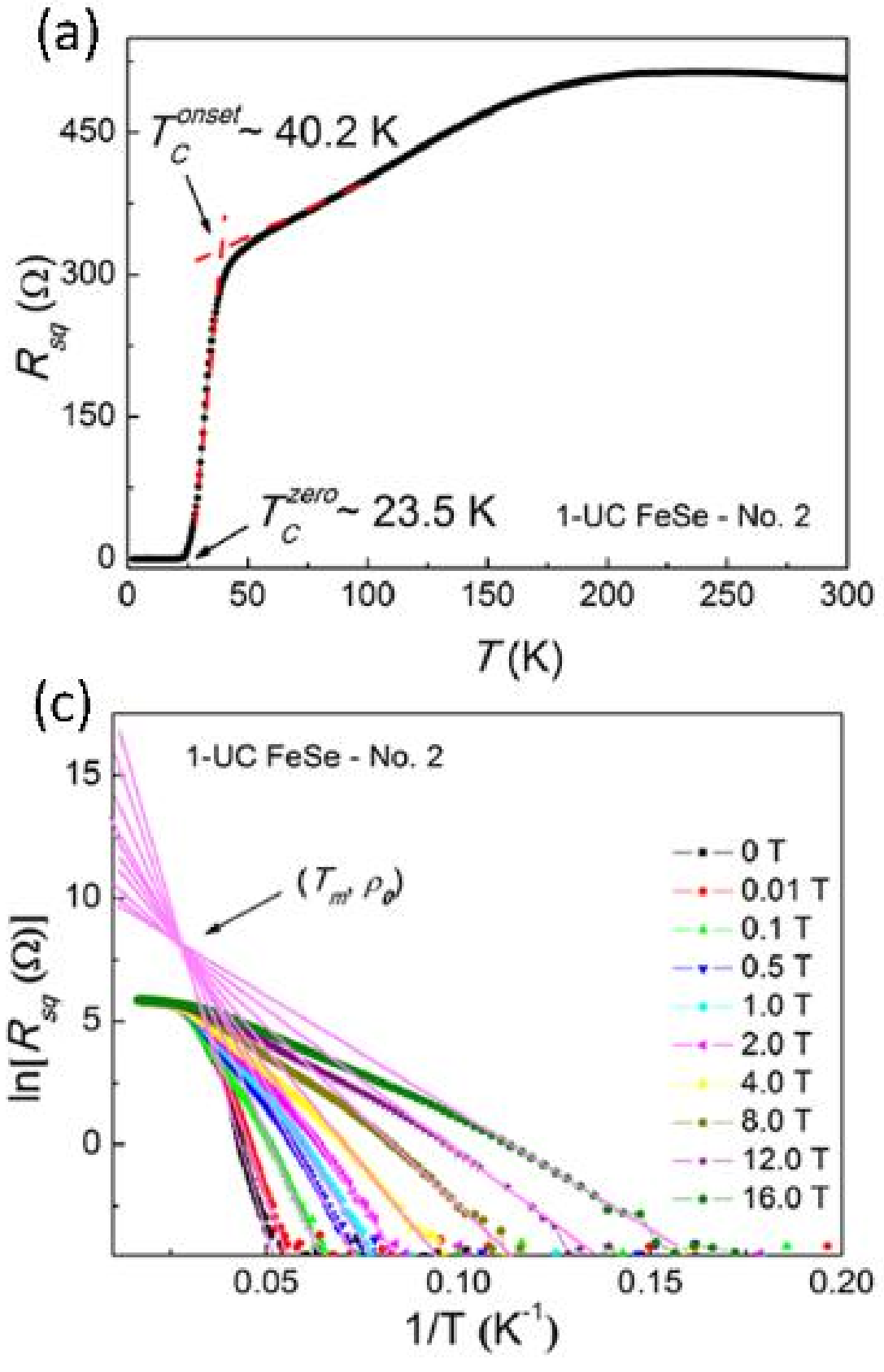

Figure 2

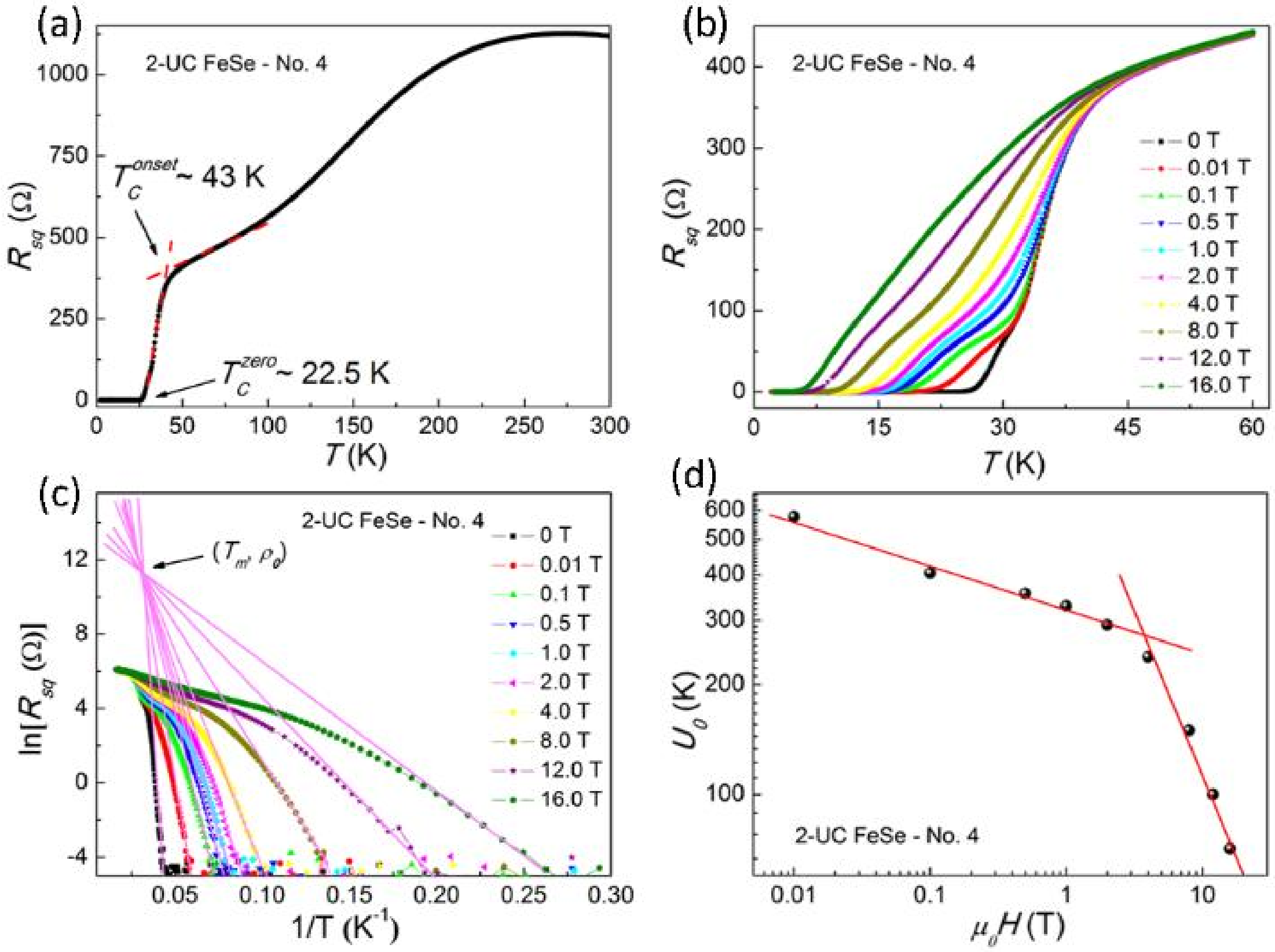

**Figure 3**

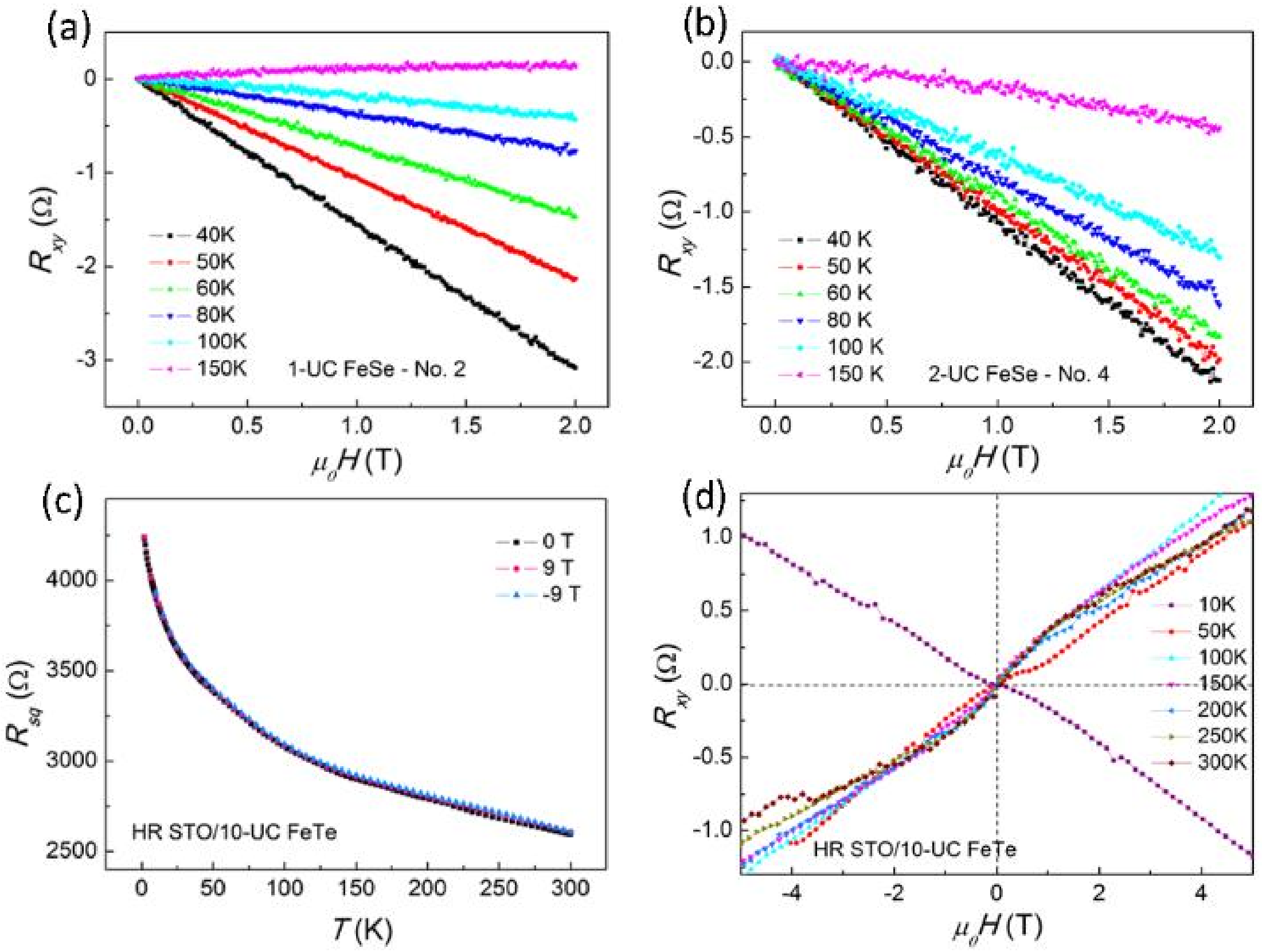

**Figure 4**

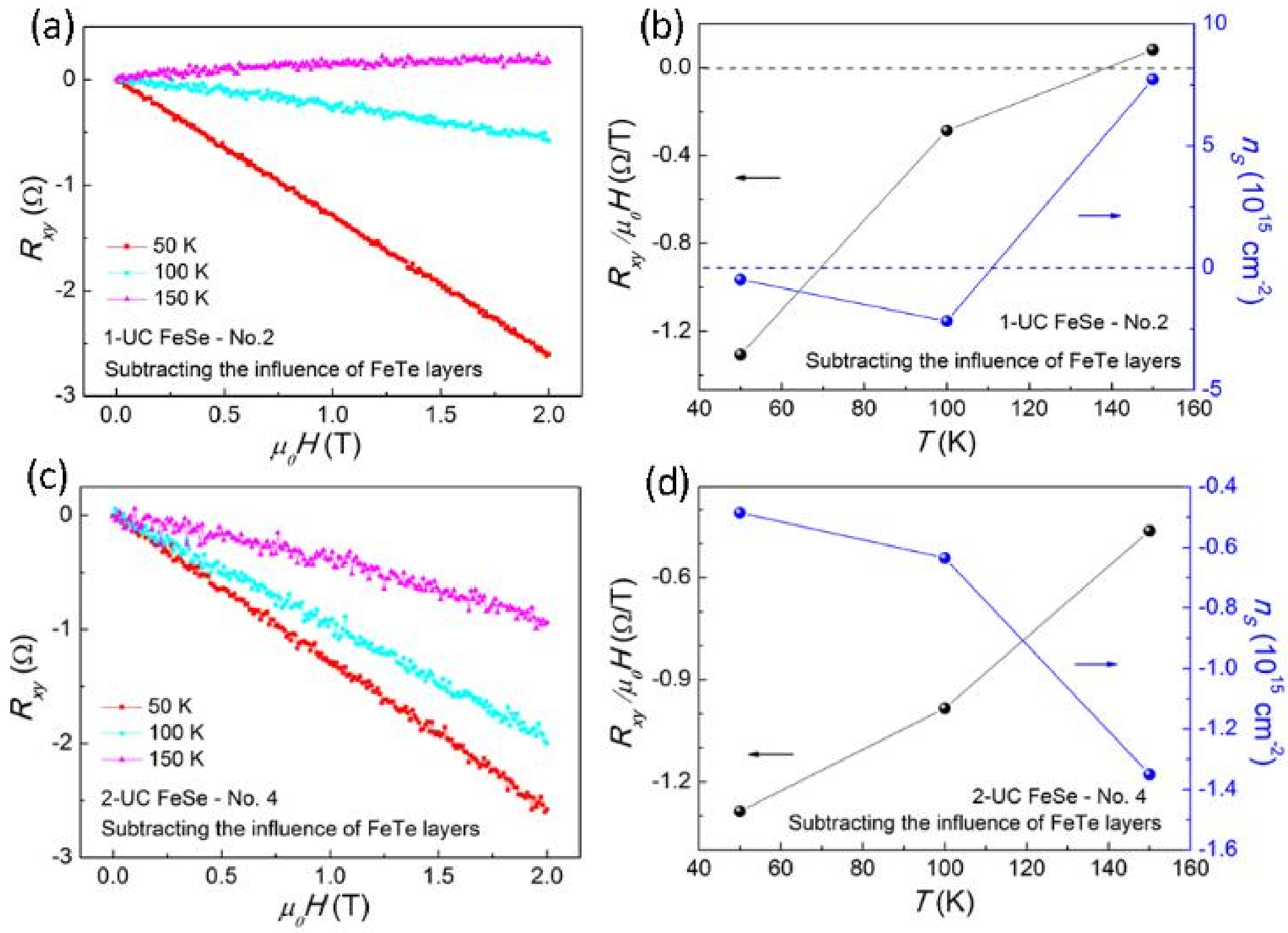

(a)
50 K
100 K
150 K
1-UC FeSe - No.2
Subtracting the influence of FeTe layers
$R_{xy}$ (Ω)
$\mu_0 H$ (T)
(b)
1-UC FeSe - No.2
Subtracting the influence of FeTe layers
$R_{xy}/\mu_0 H$ (Ω/T)
$n_s$ ($10^{15}$ cm$^{-2}$)
$T$ (K)
(c)
50 K
100 K
150 K
2-UC FeSe - No. 4
Subtracting the influence of FeTe layers
$R_{xy}$ (Ω)
$\mu_0 H$ (T)
(d)
2-UC FeSe - No. 4
Subtracting the influence of FeTe layers
$R_{xy}/\mu_0 H$ (Ω/T)
$n_s$ ($10^{15}$ cm$^{-2}$)
$T$ (K)

**Figure 5**

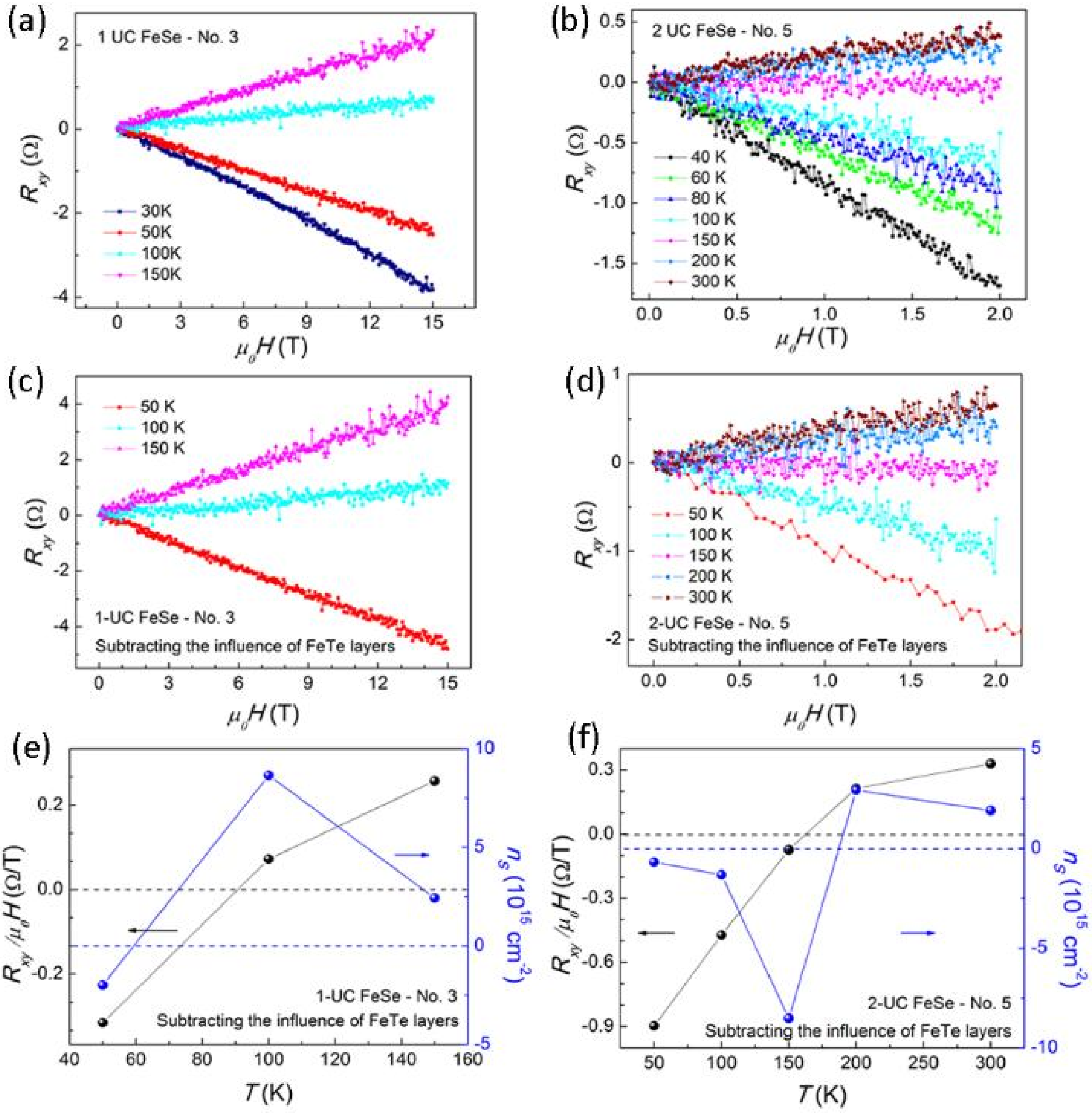
(a)
1 UC FeSe - No. 3
30K
50K
100K
150K
$R_{xy}$ (Ω)
$\mu_0 H$ (T)
(b)
2 UC FeSe - No. 5
40 K
60 K
80 K
100 K
150 K
200 K
300 K
$R_{xy}$ (Ω)
$\mu_0 H$ (T)
(c)
50 K
100 K
150 K
1-UC FeSe - No. 3
Subtracting the influence of FeTe layers
$R_{xy}$ (Ω)
$\mu_0 H$ (T)
(d)
50 K
100 K
150 K
200 K
300 K
2-UC FeSe - No. 5
Subtracting the influence of FeTe layers
$R_{xy}$ (Ω)
$\mu_0 H$ (T)
(e)
1-UC FeSe - No. 3
Subtracting the influence of FeTe layers
$R_{xy}/\mu_0 H$ (Ω/T)
$n_s$ ($10^{15}$ cm$^{-2}$)
$T$ (K)
(f)
2-UC FeSe - No. 5
Subtracting the influence of FeTe layers
$R_{xy}/\mu_0 H$ (Ω/T)
$n_s$ ($10^{15}$ cm$^{-2}$)
$T$ (K)

**Figure 6**